\documentclass[twocolumn,showpacs,reprint,preprintnumbers,amsmath,amssymb]{revtex4-1}
\usepackage{amsmath}
\usepackage{txfonts}
\usepackage{graphicx}
\usepackage{dcolumn}
\usepackage{bm}
\usepackage{color}
\usepackage[colorlinks=true,citecolor=blue,linkcolor=magenta]{hyperref}

\begin{document}

\title{Highly sensitive optical sensor for precision measurement of electrical charges based on optomechanically induced difference-sideband generation}

\author{Hao Xiong}\email{haoxiong1217@gmail.com}
\author{Zeng-Xing Liu}
\author{Ying Wu}
\affiliation{School of Physics, Huazhong University of Science and Technology, Wuhan, 430074, P. R. China}
\date{\today}

\begin{abstract}
Difference-sideband generation in an optomechanical system coupled to a charged object is investigated beyond the conventional linearized description of optomechanical interactions. An exponential decay law for difference-sideband generation in the presence of electric interaction is identified which exhibits more sensitivity to electrical charges than the conventional linearized effects. Using exact the same parameters with previous work based on the linearized dynamics of the optomechanical interactions, we show that optomechanically induced difference-sideband generation may enable an all-optical sensor for precision measurement of electrical charges with higher precision and lower power. The proposed mechanism is especially suited for on-chip optomechanical devices, where nonlinear optomechanical interaction in the weak coupling regime is within current experimental reach.

\end{abstract}

\maketitle

Micromechanical resonators, in combination with a high Q optical cavity via resonantly enhanced feedback-backaction arising from radiation pressure, can be used to manipulate light propagation exotically \cite{rev,nonreciprocity} and provides a special platform for performing precision measurement \cite{measurement1,measurement2,measurement3} and force sensors \cite{force,force2} due to their important properties of small masses and high integrability. The force sensors based on the optomechanical interaction is usually carried out via the correlations between the measured quantities and output spectra, and precision measurement of electrical charges \cite{Zhang} in an optomechanical system has been suggested based on the effect of optomechanically induced transparency, where sharp transmission features controlled by the control laser beam exhibit Coulomb-interaction dependent effect that can be well understood through the linearization of the optomechanical interactions \cite{omit1,omit2,omit3}. Compared with traditional methods, optomechanical sensors allows for remote sensing via optical fibers and relies free on naturally occurring resonances \cite{measurement1,measurement2}.

Recently, due to the prominent applications in precision measurement and optical combs, nonlinear optomechanical interactions have emerged as a new frontier in cavity optomechanics \cite{om}, and have enabled many interesting topics, such as second-order sideband generation \cite{hxsecond,delaying,sd,nonlinearcoupling}, sideband comb \cite{hxcomb,heom}, optomechanical chaos \cite{chaos}, and carrier-envelope phase-dependent effects \cite{hxhsg}. It has been shown that nonlinear signals in the optomechanical system could be a sensitive tool for performing precision measurement of the average phonon number and may provide measurement with higher precision \cite{nonlinear,nonlinear2}.

Nonlinear features of optomechanical systems with multiple probe field driven have been discussed recently \cite{hx3}, and spectral signals at difference sideband has been demonstrated analytically which provides an effective way for light manipulation and precision measurement in a solid-state architecture \cite{hx4}. In the present work, difference-sideband generation in an optomechanical system coupled to a charged object is analytically investigated and precision measurement of electrical charges by means of the signals at difference sideband is carefully examined. We identify an exponential decay law for both upper and lower difference-sideband generation which can describe the dependence of the intensities of these signals on the charge number. This exponential decay law for difference-sideband generation enables an attractive device for the measurement of electrical charges with higher precision and lower power than the conventional linearized optomechanical interaction. The effect of electrical-charge dependent difference-sideband generation is especially suited for on-chip optomechanical devices, where nonlinear optomechanical interaction in the weak coupling regime is within current experimental reach.

To relate to previous works, we emphasize that the present work can be seen as an extension of two previous papers \cite{kong,hx4}. In Ref. \cite{kong} the non-perturbative behavior (the parameters are chosen in the unstable region) of the optomechanical resonator under the influence of electro-static force was studied in detail. In \cite{hx4} the difference sideband generation in the presence of two probe fields was proposed and analyzed. Here we combine the analysis from these two papers and study the impact of the electrostatic force on difference sideband generation.

\begin{figure}[ht]
\centering
\includegraphics[width=0.40\textwidth]{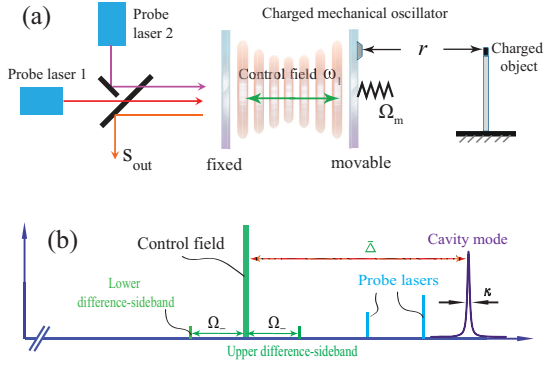}
\caption{\label{fig:1}(Color online) (a) Schematic diagram of a double probe fields driven optomechanical system where the mechanical oscillator is coupled to an adjoining charged body via coulomb force. (b) Frequency spectrogram of cavity fields in the optomechanical system. The frequency of the control field is detuned by $\bar{\Delta}$ from the cavity mode. There are difference-sideband generation due to the nonlinear optomechanical interactions.}
\end{figure}

It has been shown theoretically \cite{Zhang} that electric interaction can be introduced to an optomechanical system (such experimental configuration has not been demonstrated yet) and a schematic diagram of such charged optomechanical system is shown in Fig. \ref{fig:1}(a) where the optomechanical system is driven by a strong control field with the frequency $\omega_c$ and two probe fields with frequencies $\omega_1$ and $\omega_2$. The Hamiltonian formulation of the optomechanical system is \cite{Zhang}: $\hat{H}=\hbar \omega_0 \hat{a}^\dag \hat{a} +{\hat{p}^2}/{2m}+{m\Omega_m^2\hat{x}^2}/{2}+\hbar G \hat{x} \hat{a}^\dag \hat{a} +\hat{H}_{\mathrm{input}}+\hat{H}_{\mathrm{elec}}$, where $\omega_0$ is the resonance frequency of the cavity, $\hat{a}$ ($\hat{a}^\dag$) is the annihilation (creation) operator of the cavity field with line width $\kappa$ in the resolved-sideband regime, $\hat{p}$ ($\hat{x}$) is the momentum (position) operator of the mechanical oscillator with angular frequency $\Omega_{m}$ and mass ${m}$, $G$ is the optomechanical coupling constant \cite{Law}.
$\hat{H}_{\mathrm{input}}=\hat{H}_{\mathrm{control}}+\hat{H}_{\mathrm{probe}}$ with $\hat{H}_{\mathrm{control}}=\mathrm{i}\hbar\sqrt{\eta\kappa} \varepsilon_c (\hat{a}^\dag e^{-\mathrm{i}\omega_c t}- \hat{a} e^{\mathrm{i}\omega_c t})$ and $\hat{H}_{\mathrm{probe}}=\mathrm{i}\hbar \sqrt{\eta\kappa}(\hat{a}^\dag\varepsilon_1 e^{-\mathrm{i}\omega_1 t}+ \hat{a}^\dag\varepsilon_2 e^{-\mathrm{i}\omega_2 t}- \mathrm{H.c.})$ where $\varepsilon_i=\sqrt{P_i/\hbar \omega_i}$ ($i$=c, 1, 2) are the amplitudes of the input fields with $P_c$ the pump power of the control field and $P_1$ ($P_2$) the power of the probe field with frequency $\omega_1$ ($\omega_2$). In the parameter configuration of optomechanically induced transparency, the frequency of the control field is detuned by $\bar{\Delta}\approx -\Omega_m$ from the cavity resonance frequency. $\hat{H}_{\mathrm{elec}}=k Q_{1} Q_{2}\hat{x}/r^2$ where ${k}$ is the electrostatic force constant, ${Q_{1}}$ and ${Q_{2}}$ are the charge of mechanical oscillator and the charged body, respectively, and ${r}$ is the distance between the mechanical oscillator and the charged body [shown in Fig. \ref{fig:1}(a)]. To describe difference-sideband generation with electric interactions more clearly, in the present work we fix the charge of the mechanical oscillator ${Q_{1}}$ and only focus on the variation of ${Q_{2}}$ which can be written as ${Q_{2}}={n} {e}$ with $e$ the elementary charge and ${n}$ the charge number.

Transforming the Hamiltonian into the rotating frame at the frequency $\omega_{c}$ based on $\hat H_{1}=\hbar \omega_{c} \hat{a}^{\dagger}\hat{a} $ and $U_{t}=e^{ {-i \hat H_{1} t}/{\hbar}}=e^{-i \omega_{1} \hat{a}^{\dagger}\hat{a} t}$ gives the following Heisenberg equations \cite{rev}:
\begin{gather}
\dot a = [{\rm{i}}(\Delta  - Gx) - \kappa /2]a + \sqrt {\eta \kappa } ({\varepsilon _c} + \varepsilon_1 e^{-\mathrm{i}\delta_1 t}+\varepsilon_2  e^{-\mathrm{i}\delta_2 t}),\label{equ:2.1}\\
m\left( {\frac{{{{\rm{d}}^2}}}{{{\rm{d}}{t^2}}} + {\Gamma _m}\frac{{\rm{d}}}{{{\rm{d}}t}} + \Omega _m^2} \right)x = -\hbar G{a^*}a - \frac{{k{Q_1}{Q_2}}}{{{r^2}}},\label{equ:2.2}
\end{gather}
where $\Delta=\omega_c-\omega_0$ is the detuning of the control field from the cavity mode, $\delta_i=\omega_i-\omega_c$ ($i$=1, 2) are the frequency difference between the $i$-th probe field and the control field, $\Gamma_m$ and $\kappa$ are the decay rates of the mechanical oscillator and the intracavity field, respectively, and all operators are reduced to their expectation values, viz. $a(t)\equiv \langle \hat{a}(t)\rangle$ and $x(t)\equiv \langle \hat{x}(t)\rangle$ \cite{omit4,1,2}. Equations (\ref{equ:2.1}) and (\ref{equ:2.2}) describe the time evolution of the optomechanical system with electric interactions. The solution of these equations can be written as $a=\bar{a}+\delta a$ and $x=\bar{x}+\delta x$, with $\bar{a}=\sqrt{\eta\kappa}\varepsilon_1/(-\mathrm{i}\bar{\Delta}+\kappa/2)$, $\bar{x}=-(\hbar G |\bar{a}|^2+\xi Q_{2})/(m\Omega_m^2)$, $\bar{\Delta}=\Delta-G \bar{x}$, and $\xi=k Q_{1}/r^2$. This system is intrinsically nonlinear and bistable behavior may occur when proper parameters are chosen, which is quite similar to multiple steady states arise in nonlinear optics \cite{well} and economic evolution \cite{non2,non3} with positive feedback. The intensity of the upper branch of the bistable curve in the system shows an instability behavior for some parametric conditions and the dynamics may be chaotic in this case \cite{chaos}.

\begin{figure}[ht]
\centering
\includegraphics[width=0.4\textwidth]{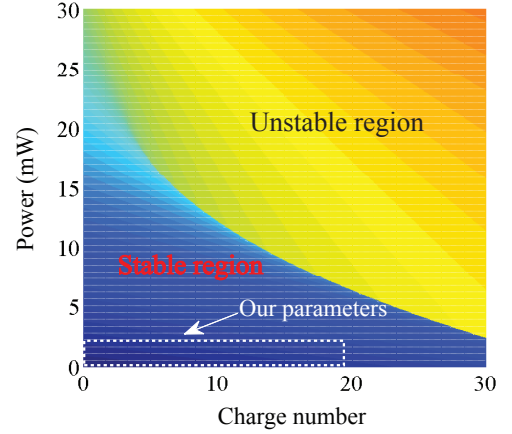}
\caption{\label{fig:2}(Color online) The parameter regime of stability in the optomechanical system with electric interactions. The parameters used in the calculation are $\Delta=\omega_{m}$, ${G/2\pi=-11}$ MHz/nm, ${m=145}$ ng, ${\kappa/2\pi= 215}$ kHz, ${\Omega_{m}/2\pi=947}$ kHz, ${\gamma_{m}/2\pi= 141}$ Hz, ${r=67 \mu}$m, and ${Q_{1}=C U}$ with ${C=27.5}$ nF and ${U=1}$ V \cite{Zhang}. The wavelength of the control field is 532 nm.}
\end{figure}

A diagram describes the parameter regime of stability is shown in Fig. \ref{fig:2} where the dynamics becomes unstable when either $P_c$ or charge number $n$ is large enough. In the present work, the power of the control field is less than 1 mW and the charge number is no more than twenty, which makes the system working in the perturbative regime, and $\delta a$ and $\delta x$ obey the following nonlinear equation:
\begin{gather}
\mho \phi  =  N{\phi ^*} + \sqrt {\eta \kappa }(\varepsilon_1 e^{-\mathrm{i}\delta_1 t}+\varepsilon_2  e^{-\mathrm{i}\delta_2 t})\sigma,\label{equ:delta_equation}
\end{gather}
where $\phi=(\delta a, \delta x)^T$, $\sigma=(1, 0)^T$, and
\begin{equation}
\mho=\left( {\begin{array}{*{20}{c}}
   {{{\rm{d}}}/{{{\rm{d}}t}} - {\rm{i}}\bar \Delta  + \kappa /2} & {{\rm{i}}G(\bar a + \delta a)}  \\
   { - \hbar G({{\bar a}^*} + \delta {a^*})} & {\hat \Psi }  \\
\end{array}} \right),
\quad
N=\hbar G \left( {\begin{array}{*{20}{c}}
   0 & 0  \\
   {\bar a} & 0  \\
\end{array}} \right),\nonumber
\end{equation}
with $\hat{\Psi}=m(\mathrm{d}^2/\mathrm{d}t^2+ \Gamma_m \mathrm{d}/\mathrm{d}t + \Omega_m^2)$. These equations of motion can be solved analytically with the linearized ansatz $\delta a^{L} =a_{\delta_1}^+e^{-\mathrm{i}\delta_1t}+a_{\delta_1}^-e^{\mathrm{i}\delta_1t} +a_{\delta_2}^+e^{-\mathrm{i}\delta_2t} +a_{\delta_2}^-e^{\mathrm{i}\delta_2t}$ and $\delta x^{L}=x_{\delta_1} e^{-\mathrm{i}\delta_1t}+x_{\delta_1}^* e^{\mathrm{i}\delta_1t}+x_{\delta_2} e^{-\mathrm{i}\delta_2t}+x_{\delta_2}^* e^{\mathrm{i}\delta_2t}$, where second- and higher- order nonlinear terms are ignored. The two spectral components at $\delta_1$ and $\delta_2$ are independent with each other in the linearized evolution, and an adjustable transparency window arises when the resonance condition is met. The linearized dynamics of the optomechanical system coupled to charged objects is studied in Ref. \cite{Zhang}. In the present work, all nonlinear terms are taken into account analytically and we focus on the effect of difference-sideband generation, which has been predicted in a traditional optomechanical system and is quite similar to difference-frequency generation in a nonlinear medium.

Following the analytical perturbation method of describing difference-sideband generation \cite{hx4}, we introduce the nonlinear ansatz of Eqs. (\ref{equ:delta_equation}): $\delta a = a_1^+e^{-\mathrm{i}\delta_1t} +a_1^-e^{\mathrm{i}\delta_1t} +a_2^+e^{-\mathrm{i}\delta_2t} +a_2^-e^{\mathrm{i}\delta_2t} +a_d^+e^{-\mathrm{i}\Omega_- t}+a_d^-e^{\mathrm{i}\Omega_- t}+\cdots$ and $\delta x=x_1 e^{-\mathrm{i}\delta_1t} +x_1^* e^{\mathrm{i}\delta_1t} +x_2 e^{-\mathrm{i}\delta_2t} +x_2^* e^{\mathrm{i}\delta_2t} +x_d e^{-\mathrm{i}\Omega_- t} +x_d^* e^{\mathrm{i}\Omega_- t}+\cdots$, with $\Omega_-=\delta_1-\delta_2$. It has been demonstrated that other frequency components, including second- and higher-order sidebands \cite{hxsecond}, can be ignored due to the fact that these components contribute little to difference-sideband generation in the perturbative regime. Substitution of the nonlinear ansatz into Eqs. (\ref{equ:delta_equation}) leads to there matrix equations \cite{hx4}: $M({\delta _1}){\alpha _1} = {\beta _1}$, $M({\delta _2}){\alpha _2} = {\beta _2}$, and $M({\Omega_-}){\alpha _d} = {\beta _d}$, where $\alpha_i=[{a_i^ + }, {{{(a_i^ - )}^*}}, {{x_i}}]^T $ with $i= 1, 2, d$, $\beta_1=[{\sqrt{\eta\kappa}{\varepsilon_1}}, 0, 0]^T$, $\beta_2=[{\sqrt {\eta\kappa}{\varepsilon_2}}, 0, 0]^T$, $\beta_d=\text{i}G[-({a_1^ + x_2^* + a_2^ - {x_1}}), {{(a_1^ - )}^*}x_2^* + {{(a_2^ + )}^*}{x_1}, {\text{i}}\hbar a_1^ + {{(a_2^ + )}^*} + {\text{i}}\hbar a_2^ - {{(a_1^ - )}^*}]^T$,
\begin{gather}
M(x) =
\left(
  \begin{array}{ccc}
   \theta (-x) & 0 & \text{i}G \bar a \\
   0 & [\theta{(x)}]^* & -\text{i}G {\bar a}^* \\
    \hbar G{{\bar a}^*} & \hbar G\bar a & {\sigma (x)}  \\
  \end{array}
\right),
\end{gather}
with $\theta (x) = s + \mathrm{i}x$, $s = \kappa /2 - {\rm{i}}\Delta  - ({\rm{i}}\hbar {G^2}|\bar a{|^2} + {\rm{i}}\xi G{Q_2})/(m\Omega _m^2)$, and $\sigma (x) = m(\Omega _m^2 - x^2 - \mathrm{i}{\Gamma _m}x)$. The first two matrix equations describe the conventional linearized optomechanical interactions, while the third equation describes the amplitude of the difference-sideband generation. The solution to these matrix equations can be obtained as follows:
\begin{gather}
a_d^ +  = \frac{{G(a_1^ + x_2^* + a_2^ - {x_1})\tau ({\Omega _ - }) - \hbar {G^2}\bar a{\xi _d}}}{{{\rm{i}}\tau ({\Omega _ - })\theta ( - {\Omega _ - }) - G(\bar xm\Omega _m^2 + \xi {Q_2})}}, \nonumber\\
a_d^ -  = \frac{{ - {\rm{i}}G \bar ax_d^* - {\rm{i}}G(a_1^ - {x_2} + a_2^ + x_1^*)}}{{\theta ({\Omega _ - })}},\label{equ:solution}
\end{gather}
where $a_i^ +  = \sqrt {\eta \kappa } {\varepsilon _i}\tau ({\delta _i})/[\theta ( - {\delta _i})\tau ({\delta _i}) + {\rm{i}}G(\bar xm\Omega _m^2 + \xi {Q_2})]$, ${x_i} =  - \hbar G{{\bar a}^*}a_i^ + /\tau ({\delta _i})$, and $a_i^ -  =  - {\rm{i}}G\bar ax_i^*/\theta ({\delta _i})$ are the amplitudes of the anti-Stoke field, mechanical oscillation, and the Stoke field, respectively, with the subscript $i= 1, 2$ denotes that the motion are driven by the $i$-th probe field. $\tau (x) = \sigma (x) + \alpha/{\theta {{(x)}^*}}$ and ${\xi _d} = a_1^ + {(a_2^ + )^*} + a_2^ - {(a_1^ - )^*} + {\rm{i}}G\bar a[{(a_1^ - )^*}x_2^* + {(a_2^ + )^*}{x_1}]/\theta {({\Omega _ - })^*}$ with $\alpha=- {\rm{i}}G(\bar xm\Omega _m^2 + \xi {Q_2})$. It can be easily verified that the solution reduces to the expression of traditional difference-sideband generation in the limit $\xi\rightarrow 0$ or $Q_2\rightarrow 0$.

\begin{figure}[ht]
\centering
\includegraphics[width=0.43\textwidth]{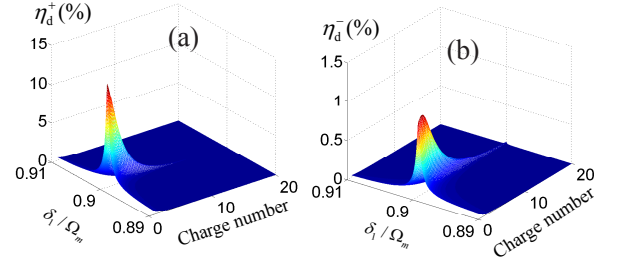}
\caption{\label{fig:3}(Color online) The efficiencies of (a) upper and (b) lower difference-sideband generation as functions of driven frequency $\Omega$ and charge number $n$. The powers of the control and probe fields are 0.1 mW and 10 $\mu$W, respectively. $\delta_2=-0.1\Omega_m$ and the other parameters are the same as Fig. \ref{fig:2}.}
\end{figure}

The amplitude of the output field at upper and lower difference sideband can be obtained as $-\sqrt{\eta\kappa}a_d^+$ and $-\sqrt{\eta\kappa}a_d^-$, respectively, based on the input-output relation $s_{\mathrm{out}}=s_{\mathrm{in}}-\sqrt{\eta \kappa }a$. Similar to Ref. \cite{hx4}, we define $\eta_d^+=|-\sqrt{\eta \kappa}a_d^+/\varepsilon_1|$ and $\eta_d^-=|-\sqrt{\eta\kappa}a_d^- /\varepsilon_1|$ as the efficiencies of the upper and lower difference-sideband generation process, where the denominator (the amplitude of the first input probe field) $\varepsilon_1$ is just chosen for convenience, and therefore leads to the efficiency being dimensionless. The efficiencies of difference-sideband generation $\eta_d^+$ and $\eta_d^-$ varies with $\delta_1$ and the charge number is shown in Fig. \ref{fig:3}. All of the parameters used in the calculation are the same as Ref. \cite{Zhang} which are chosen from the recent experiment in the resolved-sideband regime. The maximal efficiency of the upper difference-sideband generation is about 13\% which can be achieved at $\delta_1=0.9\Omega_m$ where the matching condition $\delta_1-\delta_2=\Omega_m$ is meet \cite{hx4}. Due to the far off-resonance nature of the lower sidebands, the maximal efficiency of lower difference-sideband generation is quite low (about 1\%) and seems hard to be detected.

A high dependence of the efficiencies of difference-sideband generation on the charge number $n$ is observed. As shown in Fig. \ref{fig:3}, the efficiencies of (both upper and lower) difference-sideband generation processes are reduced significantly when the electric interaction is imposed on the optomechanical system, and the higher the charge number, the weaker the signals of difference-sideband generation. The efficiency of the upper difference-sideband generation is reduced to about 4.1\% for only four charges are imposed on the charged body. These results implies that optomechanically induced difference-sideband generation can be substantively modified by electric interactions, which results in tunable optical nonlinearity and convenient optomechanical control.

The solution (\ref{equ:solution}) is made up of two terms: a term describes the process of difference-sideband generation from the downconverted probe fields and the another term arises from the process of upconverted control field. Solution (\ref{equ:solution}) shows explicitly a high dependence of the efficiencies of difference-sideband generation on the electrical charges $Q_2$, where the quantity of electric charge $Q_2$ plays an important role in both processes of difference-sideband generation from the downconverted probe fields and the upconverted control field. In the concerned resolved-sideband regime, the sidebands that are far off-resonance can be neglected (see e.g. the supporting material of Ref. \cite{omit4}), and the solution of $a_d^+$ can be simplified as
\begin{gather}
a_d^ +  \approx \frac{{ - \hbar {G^2}\bar aa_1^ + {{(a_2^ + )}^*}}}{{{\rm{i}}\tau ({\Omega _ - })\theta ( - {\Omega _ - }) + \hbar {G^2}|\bar a{|^2}}},
\end{gather}
which leads to the equation ${\rm{d}}a_d^ + /{\rm{d}}{Q_2} = -\gamma_+ a_d^+$, where the higher order terms of $a_d^+$ are ignored due to the low coefficients. Then the solution of $a_d^+$ can be expressed as $a_d^ +  = a_d^{ + (0)} \exp(-\gamma_+ Q_2)$ and similarly $a_d^ -  = a_d^{ - (0)} \exp(-\gamma_- Q_2)$, where $\gamma_\pm$ are almost independent of the electrical charges $Q_2$ in the limit $G\bar{x}\ll\Omega_m$ with $a_d^{\pm (0)}$ the solution of conventional difference-sideband generation without electric interactions.

\begin{figure}[ht]
\centering
\includegraphics[width=0.33\textwidth]{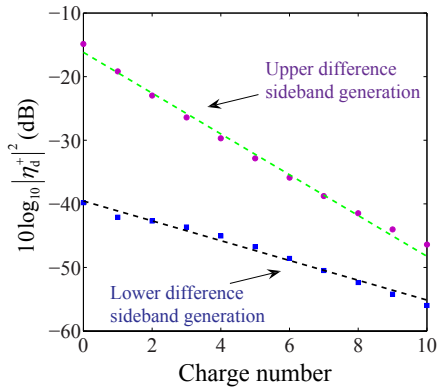}
\caption{\label{fig:4}(Color online) Dependence of the efficiency of difference-sideband generation on the charge number $n$. The parameters are the same as Fig. \ref{fig:2}.}
\end{figure}

Such exponential decay law for difference-sideband generation in the presence of electric interactions is quite accurate when the charge number is not big enough. Figure \ref{fig:4} shows the dependence of the efficiencies of difference-sideband generation (in dB form) on the charge number $n$. The linear relation between ${\log _{10}}|\eta _d^ \pm|^2$ and the charge number $n$ confirms the exponential decay law, which holds for both upper and lower difference-sideband generation.

This exponential decay law for difference-sideband generation (especially the upper process) suggests that this approach may be used for high resolution charge detection, given it shows performance metrics orders of magnitude better than previous work based on the linearized dynamics of the optomechanical interactions. Although nonlinear effects should be much weaker than the linear counterpart in general, this approach may allow measurement with power lower than the mechanism of optomechanically induced transparency, because measurement based on the optomechanically induced transparency exhibits a threshold value that the pump power must exceed it to work. Difference-sideband generation, by contrast, has no such restriction. The proposed mechanism is especially suited for optomechanical devices, where nonlinear optomechanical interaction in the weak coupling regime is within current experimental reach \cite{exp}. However, due to the exponential decay law, difference-sideband signals become very weak when $Q_2=10$ (corresponds to about $10 \sim 10^2$ photons), further noise analysis and other practical consideration are required to move in that direction.

In summary, an exponential decay law for difference-sideband generation in a hybrid optomechanical system with electric interaction is identified, which may enable an attractive device for the measurement of electrical charges with high precision. Using exact the same parameters with previous work based on the linearized dynamics of the optomechanical interactions, the optical sensor device based on difference-sideband generation shows performance metrics orders of magnitude better than the linearized case due to the sensitivity of nonlinear optomechanical interaction.

The work was supported by the NSF of China (Grant Nos. 11405061, 11774113, 11375067, and 11574104).

\end{document}